\documentclass{ws-ijmpa}
\usepackage[super,compress]{cite}
\usepackage{graphicx}
\usepackage{subfigure}
\usepackage{float}
\usepackage{xcolor}
\usepackage{multirow}
\usepackage{lscape}
\usepackage{tensor}
\usepackage{hyperref}
\usepackage{lscape}
\begin{document}
\markboth{Carrillo-Monteverde et.al}{PHENOMENOLOGY OF THE PA2HDM WITH LEPTONIC MESON DECAYS}
\catchline{}{}{}{}{}

\bibliographystyle{ws-ijmpa}
%

%

\title{\bf PHENOMENOLOGY OF THE PARTIALLY ALIGNED 2HDM WITH LEPTONIC MESON DECAYS}

\author{A. Carrillo-Monteverde}

\address{Department of Physics and Astronomy, University
    of Sussex, Brighton BN1 9QH, UK\\
A.carrillo-Monteverde@sussex.ac.uk}

\author{S. G\'omez-\'Avila}

\address{\'Area Acad\'emica de Matem\'aticas y F\'isica,
    Universidad Aut\'onoma del Estado de Hidalgo,\\
    Carr. Pachuca-Tulancingo Km. 4.5, C.P. 42184, Pachuca, Hidalgo,
    Mexico.\\selim\_gomez@uaeh.edu.mx}

\author{R. G\'omez-Rosas}
\address{Instituto de F\'isica, BUAP, Apdo. Postal J-48,
    C.P. 72570 Puebla, Pue., Mexico}

\author{L. Lopez-Lozano\footnote{Corresponding author.}}
\address{\'Area Acad\'emica de Matem\'aticas y F\'isica,
    Universidad Aut\'onoma del Estado de Hidalgo,\\
    Carr. Pachuca-Tulancingo Km. 4.5, C.P. 42184, Pachuca, Hidalgo,
    Mexico.\\
lao\_lopez@uaeh.edu.mx}
  
  \author{A. Rosado}
\address{Instituto de F\'isica, BUAP, Apdo. Postal J-48,
  C.P. 72570 Puebla, Pue., Mexico\\
rosado@ifuap.buap.mx}

\maketitle 

\begin{history}
\received{Day Month Year}
\revised{Day Month Year}
\end{history}

\begin{abstract}
  In this paper we present a phenomenological analysis of the
  Partially Aligned Two Higgs Doublet Model (PA-2HDM) by using
  leptonic decays of mesons and $B^0_{d,s}$-$\bar B^0_{d,s}$ mixing. We focus our attention in a scenario where the leading contribution to FCNC is given by the tree level interaction with the light pseudoscalar $A^0$ ($M_{A^0}\sim 250$ GeV).  We show how an underlying flavor symmetry controls FCNC in the quark and lepton couplings with the pseudoscalar, without alignment between Yukawa matrices.  Upper bounds on the free parameters are calculated in the context of the leptonic decays $B^0_{s,d}\to\mu^+\mu^-$ and $K^0_L\to \mu^+\mu^-$ and $B^0_{s,d}$ mixing. Also, our assumptions implies that bounds on New Physics contribution in the quark sector coming from $B^0_{s,d}$ mixing impose an upper bound on the parameters for the leptonic sector. Finally we give predictions of branching ratios for leptonic decay of mesons with FCNC and LFV.

\end{abstract}

\ccode{PACS numbers:11.30.Hv, 12.15.Ff, 13.20.-v, 14.80Cp }

\section{Introduction}
Up to the present day, no experimental result in particle physics has
significantly deviated from the theoretical predictions of the
Standard Model (SM)\cite{Olive2016}. However, there is a widespread
consensus on the need for a more fundamental theory to address the
open questions left unexplained by the SM. Despite some small signs of
non-universality in leptonic gauge couplings and some tensions in the
measurements of meson decays, to mention some examples, there is a
pronounced lack of information guiding the search for new physics that
can address open questions like the flavor problem or the nature of
dark matter.

In this uncertain view, the Two Higgs Doublet Model (2HDM)
\cite{Lee1973,Branco2012} plays a role as a minimal extension of the
SM and a suitable way to parameterize new physics. By itself, this
model does not solve the open problems of the Standard Model but it
can be thought as an intermediate framework in the search of more
fundamental explanations. Without new symmetries, this model introduces Flavor Changing Neutral Currents (FCNC) at tree level, in contrast with the SM where they are absent. The different versions of the 2HDM are determined by the way FCNC are suppressed.

The 2HDM type I \cite{Haber1979,Hall1981} and 2HDM type II
\cite{Donoghue1979} versions impose discrete symmetries $Z_2$ in the scalar
sector with the goal of eliminating FCNC at tree level, generating models with Natural Flavor Conservation (NFC)\cite{Glashow1977}.

In other versions of the 2HDM, where small FCNC appear at tree level, another
suppression mechanism is needed. This occurs, for example, in the 2HDM
type III (2HDM-III) \cite{Atwood1997a, Diaz-Cruz2005, Diaz-Cruz2009,
  Mahmoudi2010} where the control of the FCNC is achieved through
constraints on the non-SM interactions that generate a hierarchy in
the matrix elements of the Yukawa matrix \cite{Froggatt1979},
or through the use of a suitable texture \cite{Fritzsch2003,Fritzsch1977} with the Cheng and Sher ansatz \cite{Cheng1987}. A Yukawa texture matrix is a term that refers to the specific way to put zeros in some entries of the matrices and the hierarchy of the non-zero values coming from flavor symmetries beyond SM.  The relation between discrete symmetries and zero-textures has been extensively studied \cite{Serodio2013}.

 Another way to eliminate FCNC is through the alignment between Yukawa matrices,
  leading to additional phenomenology of the 2HDM
  \cite{Zhou2004,Pich2009}. Aligment of Yukawa matrices is a sufficient condition to diagonalize mass matrix of fermions eliminating tree level FCNC. As previous works have shown, the alignment is not necessarily stable under renormalization group flow
  \cite{Ferreira2010a}. Also alignment is not a necessary condition, but a less restrictive condition: the products $Y_1^fY_2^{f\dagger}$ and $Y^f_2Y_1^{f\dagger}$ should be normal matrices \cite{Das2018a}.

  Many analysis as in references \citen{Aoki2009,Xiao2004,Becirevic2002,Aoki2019,Urban1998,Barger1990}, argued that scalar extensions of the SM cannot contain  FCNC at tree level in order to fullfill the restrictions imposed by $K^0-\bar K^0$ and $B^0_d-\bar B^0_d$ mixing. Experimental results show a noticeable agreement with SM, leaving only a small window for NP contributions without new assumptions. As it was pointed out by Xiao and Guo \cite{Xiao2004} and others, such suppression makes reasonable to assume that Yukawa couplings with fermions involving the first generation are zero. As a consequence, the main contribution to neutral meson mixing comes from box and penguin diagrams with charged scalars, as well as for the decays $B_{s,d}^0\to \mu^+\mu^-$ and $K^0_L\to\mu^+\mu^-$.
  
Nevertheless, in an era of precision measurements in B factories, we consider these assumptions too restrictive. Instead, we perform a phenomenological analysis with updated experimental results and lattice predictions on decay constants of meson in order to estimate the order of magnitude of the Yukawa couplings with a light pseudoscalar. Also, to complete our analysis, we explore masses of the order of TeV.
  
In this paper, we take a look at the constraints suggested by flavor phenomenology, in particular those coming from flavor violating processes, in the context of the Partially Aligned 2HDM scenario (PA-2HDM)\cite{Hernandez-Sanchez2012}. This analysis parameterizes the non-standard Yukawa matrix elements using the Cheng and Sher ansatz in a generic sense, in order to ease comparison with the literature where the 2HDM-III has been studied. In particular, we choose a parametrization where the coupling of the scalars and fermions takes the form of a modified 2HDM-II but with non diagonal contributions.

We analize the impact on the Yukawa matrix elements coming by setting to zero the elements related with the couplings of the pseudoscalar with fermions of one generation per time. These zeros come from a specific flavor transformation on the mass matrix, that corresponds to a subgroup of $SU(3)$. The subgroups considered  here are the Iso-Spin, V-Spin and U-Spin. Also, we assume that the same parametrization for quarks is valid for leptons and the only difference is regulated by the hierarchy of masses. This suggest the existence of a Universal Texture similar to those mentioned by Y. Koide (see for instance \citen{Koide2004} and references therein), but in a different level. This assumption allows to relate bounds on the quark sector with bounds in the leptonic sector.
  
We consider textures with pairwise fermion family mixing. Similar textures are often associated with discrete symmetries, like the
nearest-neighbour interaction (NNI) texture \cite{Branco2010} obtained
from a $Z_4$ flavor group.

Instead of assuming only the case where the couplings with the first generation are zero \cite{Xiao2004,Barger1990,Atwood1997a}, we analize the three possibilities to bound the couplings and the mass of the pseudoscalar $A^0$ coming from the 2HDM. In this sense, the suppresion of FCNC in the leptonic decay of mesons is the consequence of three aspects: \emph{(i)} a symmetry in the potential, \emph{(ii)} the dependence of couplings in terms of the hierachy of masses (a feature that comes from the PA-2HDM) and \emph{(iii)} the restrictions from  $B^0_s-\bar B_s^0$ and $B^0_d-\bar B^0_d$ mixing. Notably, the symmetry that set some Yukawa elements to zero is  lifted to a continuous $U(1)$ symmetry when the scalar potential is  included; as it was shown by Ivanov\cite{Ivanov2013}, the smallest setup  that can include a $Z_4$ symmetry is a Three Higgs Doublet Model (3HDM). In principle, this may be inconvenient because a light Goldstone boson will appear in the spontaneous symmetry breaking (SSB) phase. In our model, we do not take into account such problems, considering the 2HDM model a special case of the 3HDM with an additional flavor symmetry $U_S(1)\times SU_F(3)$ that decoupled the third doublet of this model.

In order to test the phenomenological viability of every subgroup of flavor transformations, we use the channels $K_L^0\to \mu^+\mu^-$, $B^0_d\to\mu^+\mu^-$ and $B^0_s\to\mu^+\mu^-$, because these are the most sensitive models with FCNC. To set the order of magnitude on the parameters of the flavor transformations, we use $B^0_s-\bar B^0_s$ and $B^0_d-\bar B^0_d$ mixing.

This paper is organized as follows; in section \ref{section2} we describe the relation between the pairwise-mixing textures of the extra Yukawa matrices and general flavor transformations in the mass basis. Also, the couplings of the PA-2HDM are deduced and we define the relevant parameters of the model. In section \ref{section3}, we set restrictions on the parameters of the flavor transformation from the phenomenological effective analysis of meson decays. In section \ref{conclusions} we give our conclusions.

\section{Flavor symmetries and the PA-2HDM}\label{section2}
The Yukawa Lagrangian in a general 2HDM is given by
\begin{eqnarray}
  -\mathcal{L}_Y&=&\bar{Q}_{Li}\left[Y^d_{a,ij}\Phi_a D_{Rj}+Y^u_{a,ij}\tilde{\Phi}_aU_{Rj}\right]\nonumber\\
                &&+\bar{L}_{Li}Y_{a,ij}^\ell\Phi_a E_{Rj}+\text{H.c.},
\end{eqnarray}
where $i,j$ are flavor indexes and  $a=1,2$. $Q_{Li}= (U_{Li},D_{Li})$
and $L_{Li}= (N_{Li},E_{Li})$ are the fermion $SU(2)$ doublets, and
$D_{Rj}$ and $U_{Rj}$ are the fermion singlets. After SSB, the mass
matrix for every sector is given by
\begin{equation}
  \label{eq:mass_matrix}
  M^f_{ij}=\frac{1}{\sqrt{2}}\left(v_1Y^f_{1,ij}+v_2Y^f_{2,ij}\right),
\end{equation}
where $f=u,d,\ell$ and $v_1$ and $v_2$ are the vacuum expectation values (VEV) of each doublet. We use (\ref{eq:mass_matrix}) to write down the Feynman rules in terms of the mass matrix $M^f_{ij}$ and $Y^f_{2,ij}$ for the couplings of down-type quarks and leptons. This allows us to write interactions as deviations from the 2HDM-II, the most studied version of the 2HDM because it represents a low energy limit of the Yukawa sector of the MSSM\cite{Gunion1992,Gunion2000}.

Without loss of generality, we can parameterize the $Y^f_{2,ij}$ contributions with a flavor transformation of the following form
\begin{equation}
  \label{eq:flavor_transformation}
  Y_{2,ij}^f=\frac{1}{v}\left(A^{f}_{2L}M^f A_{2R}^{f}\right)_{ij}.
\end{equation}
where $v^2=v_1^2+v_2^2$ and $A_{2L(R)}^f$ are $SU(3)$
matrices \cite{Lopez-Lozano2016a}.

We want to point out that in the PA-2HDM \cite{Hernandez-Sanchez2012} the mechanism to suppress FCNC comes from flavor symmetries introduced by $A_{2L(R)}$ matrices. In this sense, we start from a generic 2HDM without prior assumptions on the Yukawa matrix elements as in the 2HDM-III, where the Cheng and Sher ansatz controls the FCNC \cite{Diaz-Cruz2005,Diaz-Cruz2009}, thus we express the Yukawa matrix elements as
\begin{equation}
  \label{eq:cheng_and_sher}
  Y^f_{2,ij}=\frac{\sqrt{m^f_im^f_j}}{v}\chi^f_{2,ij},\quad\text{for}\quad f=d,\ell
\end{equation}
where the parameter $\chi^f_{2,ij}$ takes any complex value. Although the expression   (\ref{eq:cheng_and_sher}) was proposed in the context of the 2HDM-III, in this work we also take the same parametrization in order to quantify the differences of this model respect to the PA-2HDM. The mass basis is defined by a set of unitary transformations on the fermion fields: $ U_{L(R)i}= V^u_{L(R),ij}u_{L(R)j}$, $D_{L(R)i} = V_{L(R),ij}^dd_{L(R)j}$ and $E_{L(R)i} = V^\ell_{L(R),ij}\ell_{L(R)j}$, that diagonalize the mass matrix (\ref{eq:mass_matrix}) to $\tilde M^f=\text{diag}\left(m^f_1,m^f_2,m^f_3\right)$.

In the mass basis, the free parameters of the 2HDM take the following form
\begin{equation}
  \label{eq:chi_ver1}
  \tilde\chi_{2,ij}^f =\frac{1}{\sqrt{m_i^fm_j^f}} \sum_{k=1}^3m_k^f(\tilde A_{2L}^f)_{ik}(\tilde A_{2R}^f)_{kj}
\end{equation}
where $\tilde\chi_{2,ij}^f=V_{L}^{f\dagger}\chi_{2,ij}^f V_{R}^{f}$ and $\tilde A_{2L(R)}^f=V_{L}^{f\dagger}A_{2L(R)}^fV_{R}^{f}$
represent the transformation from the flavor basis to the mass basis. The expression (\ref{eq:chi_ver1}) allows us to explore different scenarios compared with the 2HDM-III and also different mechanisms to control FCNC.

From equation (\ref{eq:chi_ver1}), we can observe that in order to avoid large FCNC coming from the heaviest fermion mass $m^f_3$, at least the elements $(\tilde A_{aL}^f)_{i3} \ll 1$; other terms might vanish if flavor symmetries are considered. Indeed, there can be several symmetries consistent with a particular texture in the Yukawa sector.

To track the suppression of the FCNC, coming from the scalar mass scale and an underlying flavor symmetry, we introduce additional hypotesis to reduce the number of free parameters to analyse the phenomenological consequences. In order to develope this study, we take the simplest approach by choosing flavor transformations (\ref{eq:flavor_transformation}) that mix only two generations at a time, thus we make the following assumptions \cite{Lopez-Lozano2016a,Hernandez-Sanchez2012}:

\begin{enumerate}
\item The Yukawa matrices are hermitians. This is always possible through weak basis transformations and it is phenomenologically viable.

\item The matrices $\tilde A^f_{2L}$ and $\tilde A^f_{2R}$ are $SU(3)$ matrices that belong to a particular subgroup, i.e. U-Spin, V-Spin or Isopin. An underlying flavor symmetry on the fermions and Higgs doublets could lead to small FCNC and in some cases to none.
  
\item In order to reduce the number of free parameters, we assume $\tilde A^f_{2L}=\tilde A^{f\dagger}_{2R}$, which means that the mass matrix transformation is unitary. For simplicity, we will omit the index $2R$ and $2L$.

\end{enumerate}

In leptonic and semileptonic decays of mesons with down-type quarks, we get 9 combinations of the subgroups of $SU(3)$. In the simplest scenario, where the same subgroup describes the leptonic and down-type Yukawa couplings, we have 8 free parameters, 4 for each sector. In order to evaluate the phenomenological viability of the PA-2HDM, we assume exactly the same transformation for every sector and then we reduce the number of free parameters to four. One of the free parameters is the complex phase of non diagonal Yukawa matrix elements; its effect on the branching ratio, coming from the pseudoscalar interaction, can be relevant when there is flavor changing at tree level for quarks and, on the other hand, conservation of leptonic flavor. Nevertheless, in the present work we will set this phase to zero.

With these conditions, we have three possibilities for the matrices $\tilde A^f$'s that mix generations. And they can be written using the Gell-Mann matrices as a basis as follows:
\begin{equation}
  \label{eq:flavor_transf}
  \tilde A^{f,X}=c^X_0+c^X\cos\theta_X\lambda_i+c^X\sin\theta_X\lambda_j+c^X_3\lambda_k,
\end{equation}
where $X=\{I,U,V\}$ and $\lambda_i=\{\lambda_1,\lambda_4,\lambda_6\}$,
$\lambda_j= \{\lambda_2, \lambda_5, \lambda_7\}$ and $\lambda_k= \{\lambda_3, \frac{1}{2}(\lambda_3 + \sqrt{3} \lambda_8), \frac{1}{2}(\lambda_3-\sqrt{3}\lambda_8)\}$ are chosen according to the subgroup that describes the mixing of fermions, the parameters $c^{I,U,V}_{0,3 }$ and $c^{I,U,V}$ are real, and $c^{I,U,V}>0$. The explicit form of the $\tilde A^{f,X}$ matrices is shown in Table \ref{tab:matrices}.

\begin{table}[ph]
  \centering
  \tbl{Here, we show the matrix elements in terms of the
    transformation parameters $c^{X}$ when we consider only one subgroup of SU$(3)$
    at a time. This leads to the mixing of two generations that
    restrict the FCNC contribution to only a small number of decays.}
{  \begin{tabular}{c c c}

    \hline\hline
    Subgroup  & $\tilde A^{f,X}$ & Basis ($\lambda_i,\lambda_j,\lambda_k$) \\
    \hline
    Iso-Spin & $\left( \begin{array}{ccc}
                         c^I_3+c^I_0  & c^Ie^{i\theta_I} & 0 \\ 
                         c^Ie^{-i\theta_I} & c^I_0-c^I_3  & 0 \\
                         0      &    0     & c^I_0
                       \end{array} \right)$ & $\lambda_1,\lambda_2, \lambda_3$\\\\
    U-Spin & $\left( \begin{array}{ccc}
                       c^U_3+c^U_0  & 0 & c^Ue^{i\theta_U} \\ 
                       0 & c^U_0  & 0 \\
                       c^Ue^{-i\theta_ U}     &    0     & c^U_0-c^U_3
                     \end{array} \right)$& $\lambda_4, \lambda_5 , \frac{1}{2}(\lambda_3+\sqrt{3}\lambda_8)$ \\\\
    V-Spin & $\left( \begin{array}{ccc}
                       c^V_0  & 0 & 0 \\ 
                       0 & c^V_0+c^V_3  & c^Ve^{i\theta_V} \\
                       0     &    c^Ve^{-i\theta_V}    & c^V_0-c^V_3
                     \end{array} \right)$&$\lambda_6, \lambda_7 , \frac{1}{2}(\lambda_3-\sqrt{3}\lambda_8)$ \\
    \hline\hline
  \end{tabular}\label{tab:matrices}}
\end{table}
 
The zeros in the Yukawa matrices can be set up by following methods widely studied \cite{Serodio2013,Grimus2004,Ferreira2011}. Textures matrices\cite{Fritzsch2003} and abelian symmetries are related via the Smith normal form. Discrete abelian symmetries of the Yukawa Lagrangian may be non minimally realized when the scalar potential is considered, and therefore lifted to continuos $U(1)$ symmetries. There are more possibilities, related to non-abelian groups, that will not be explored in this work.
  
With the above assumptions and considering that the Yukawa matrices are hermitian, i.e. $\tilde\chi_{2,ji}^f= (\tilde\chi_{2,ij}^f)^*$, we can obtain the explicit expressions for $\tilde\chi$'s shown below.

For the Iso-Spin texture, we obtain the equations
\begin{eqnarray}
 \label{eq:ISpinChi}
  \tilde\chi^{f}_{2,11}&=& (c^I_0+c^I_3)^2+\frac{m^f_2}{m_1^f}(c^I)^2\\
  \tilde\chi^{f}_{2,22}&=&(c^I_0-c^I_3)^2+\frac{m^f_1}{m_2^f}(c^I)^2\\
  \tilde\chi^{f}_{2,33}&=&(c^I_0)^2\\
\tilde\chi^{f}_{2,12}&=& c^{I}e^{i\theta_{I}}\left[\sqrt{\frac{m_1^f}{m_2^f}} (c^I_0+c^I_3)+\sqrt{\frac{m^f_2}{m^f_1}}(c^I_0-c^I_3)\right] 
\end{eqnarray}
and $\tilde\chi_{2,13}^{f}=\tilde\chi_{2,23}^{f}=0$.

For the U-Spin texture, the expressions are given by
\begin{eqnarray}
  \label{eq:USpinChi}
    \tilde\chi^{f}_{2,11}&=&(c^U_0+c^U_3)^2+\frac{m^f_3}{m_1^f}(c^U)^2  \\
  \tilde\chi^{f}_{2,22}&=&(c^U_0)^2\\
  \tilde\chi^{f}_{2,33}&=&(c^U_0- c^U_3)^2+\frac{m^f_1}{m_3^f}(c^U)^2\\
\tilde\chi^{f}_{2,13}&=&c^Ue^{i\theta_{U}}\left[\sqrt{\frac{m^f_1}{m_3^f}} (c^U_0+c^U_3) + \sqrt{\frac{m^f_3}{m^f_1}}(c^U_0-c^U_3)\right]
\end{eqnarray}
with $\tilde\chi^{f}_{2,23} = \tilde\chi_{2,12}^{f}=0$.

And for the V-Spin texture, the parameters are given by
\begin{eqnarray}
    \tilde\chi^{f}_{2,11}&=&(c^V_0)^2 \\
  \tilde\chi^{f}_{2,22}&=&(c^V_0+ c^V_3)^2+\frac{m^f_3}{m_2^f}(c^V)^2 \\
  \tilde\chi^{f}_{2,33}&=&(c^V_0- c^V_3)^2+\frac{m^f_2}{m_3^f}(c^V)^2\\
\tilde\chi^{f}_{2,23}&=&c^Ve^{i\theta_V}\left[\sqrt{\frac{m^f_2}{m_3^f}}(c^V_0+c^V_3) + \sqrt{\frac{m^f_3}{m^f_2}}(c^V_0-c^V_3)\right]\label{eq:VSpinChi}
\end{eqnarray}
with $\tilde\chi_{2,13}^{f}=\tilde\chi_{2,12}^{f}=0$.

In cases with highly hierarchical fermion masses and in order to control FCNC, it is expected that $c^{I,U,V}_0\simeq c^{I,U,V}_3$. As the masses of quarks and charged leptons are highly hierarchical, the relation $c_0^{I,U,V}=c_3^{I,U,V}$ suppresses the problematic term that appears in the $\chi$'s expressions with the factor $\sqrt{\frac{m^f_i}{m_{j}^f}}$ where $m^f_{i}>>m_{j}^f$. However, we will keep these contributions in our numerical analysis. For the neutrino sector in this parametrization, $c_0^{I,U,V}$ and $c_3^{I,U,V}$ could be very different because of their unknown mass hierarchy.

\section{Phenomenology of the Leptonic Meson decays in the PA-2HDM}\label{section3}
In order to perform a phenomenological analysis, we use leptonic decay of pseudoscalar mesons. In the PA-2HDM, the spontaneous CP symmetry is absent and the VEVs $v_1$ and $v_2$ are real. At tree level, the only non-standard contribution to processes of the form $P^0\to\ell^+\ell^-$, where $P^0$ is a neutral pseudoscalar meson ($B^0_{s,d},K^0_L$) and $\ell^\pm$ a charged lepton ($e,\mu$), comes from the interchange of the pseudoscalar $A^0$, reducing the number of free parameters to the ratio of the VEVs, $\tan\beta=v_2/v_1$, the mass of the pseudoscalar $M_A$ and the non standard couplings $\tilde\chi_{2,ij}^{f}$ ($f=d,\ell$). Note that at low energies (when the $A^0$ is intregrated out to get an effective Hamiltonian), this pseudoscalar interaction might represent a parametrization of a general pseudoscalar coupling where $M_A$ is the scale of NP contributions to the leptonic decays of mesons. Although charged scalars contribute only at the one loop level \cite{Logan2000a}, if we assume that the Yukawa parameters and masses of all scalars are similar, their one loop contribution is very small compared with tree-level neutral currents. In scenarios where the masses of charged and neutral scalars are very different, we need to take into account higher-order diagrams.

In general, the Wilson operators that contribute to leptonic decays of pseudoscalar mesons at low energies are \cite{Buchalla1996}:
\begin{equation}
  \label{eqn:operadores}
  \begin{split}
    \mathcal{O}_V^{q_fq_i} &=\left(\bar{q}_f\gamma_{\mu}P_Lq_i\right)\left(\bar{l}_B\gamma^{\mu}l_A\right), \\
    \mathcal{O}_A^{q_fq_i} &=\left(\bar{q}_f\gamma_{\mu}P_Lq_i\right)\left(\bar{l}_B\gamma^{\mu}\gamma_5l_A\right),\\
    \mathcal{O^{'}}_V^{q_fq_i}&= \left(\bar{q}_f\gamma_{\mu}P_Rq_i\right)\left(\bar{l}_B\gamma^{\mu}l_A\right), \\
    \mathcal{O^{'}}_A^{q_fq_i}&= \left(\bar{q}_f\gamma_{\mu}P_Rq_i\right)\left(\bar{l}_B\gamma^{\mu}\gamma_5l_A\right),\\
    \mathcal{O}_S^{q_fq_i}&= \left(\bar{q}_f P_Lq_i\right)\left(\bar{l}_Bl_A\right), \\
    \mathcal{O}_P^{q_fq_i}&= \left(\bar{q}_fP_Lq_i\right)\left(\bar{l}_B\gamma_5l_A\right),\\
    \mathcal{O^{'}}_S^{q_fq_i}&= \left(\bar{q}_f P_Rq_i\right)\left(\bar{l}_Bl_A\right), \\
    \mathcal{O^{'}}_P^{q_fq_i}&= \left(\bar{q}_fP_Rq_i\right)\left(\bar{l}_B\gamma_5l_A\right).
  \end{split}
\end{equation}
And the effective Hamiltonian can be written as
\begin{equation}
  \label{eqn:Heffc}
  \begin{split}
    \mathcal{H}_{eff}=-\frac{G_F^2M_W^2}{\pi^2}[C_V^{q_fq_i}O_V^{q_fq_i}+C_A^{q_fq_i}O_A^{q_fq_i}+C_S^{q_fq_i}O_S^{q_fq_i}+C_P^{q_fq_i}O_P^{q_fq_i}\\+
    \text{primed}] + H.c..
  \end{split}
\end{equation}
When we restrict our analysis to SM and NP tree level pseudoscalar contributions, only axial and pseudoscalar Wilson coefficients are required. Excluding box and penguin diagrams with off-shell charged scalars, the branching ratio takes the following form \cite{Crivellin2013}
\begin{equation}
  \label{eqn:BR^{Exp}}
  \begin{split}
    \mathcal{B}[P^0_{\bar{q}_f,q_i}\to
    l_A^+l_B^-]&=\frac{G_F^4M_W^4}{32\pi^5}f(x_A^2,x_B^2)M_{P^0}f_{P^0}^2(m_{l_A}+m_{l_B})^2\tau_{P^0}
    [1-(x_A-x_B)^2] \\ & \left| \frac{M_{P^0}^2\left(
          C_P^{q_fq_i}-C_p^{'q_fq_i}\right)}{\left(m_{q_f}+m_{q_i}\right)\left(m_{l_A} + m_{l_B}\right)}-\left(C_A^{q_fq_i}-C_A^{'q_fq_i}
      \right)\right|^2,
  \end{split}
\end{equation}  
where $f(x_i,x_j)$ is defined by:
\begin{eqnarray}
  f(x_i,x_j)=\sqrt{1-2\left(x_i+x_j\right)+\left(x_i-x_j\right)^2},
\end{eqnarray}
with $x_i=\frac{m_{l_i}}{M_{P^0}}$.
 The SM contributions to the branching ratio are contained in the Wilson coefficient $C_A$   \cite{Crivellin2013}. This calculation is reported in the literature for the pseudoscalar meson $B$ \cite{Bobeth2014c,Bobeth2014b,Bobeth2001b}.

The couplings with the pseudoscalar $A^0$ are given using the parametrization in Diaz et al.\cite{Diaz-Cruz2005}
\begin{equation}
  \label{couplingsAqiqf}
  \Gamma_{A^0q_fq_i}=\frac{i }{2}\left[-m_i \tan\beta \delta_{fi}+\frac{\sqrt{m_{q_i}m_{q_f}}}{\sqrt{2}\cos\beta} \tilde \chi_{2,q_fq_i}^d\right],
\end{equation}
\begin{equation}
  \label{couplingsAlalb}
  \Gamma_{A^0\ell_A \ell_B}=\frac{i}{2}\left[-{m_\ell} \tan\beta \delta_{AB}+ \frac{ \sqrt{m_{\ell_A} m_{\ell_B}} }{\sqrt{2}\cos\beta} \tilde\chi^\ell_{2,\ell_A\ell_B}\right],
\end{equation}
where the first term in each equation do not contribute to the FCNC and corresponds to the modification of the 2HDM-II over the SM; the second term contains flavor violation and it is proportional to the second Yukawa matrix as a consequence of the chosen parametrization. It is easy to deduce that at low energies
\begin{equation}\label{Wilson_PCoeffcient}
  C_P-C'_P=\frac{-i}{m_A^2}\Gamma_{A^0q_iq_f}\Gamma_{A^0\ell_A \ell_B}.
\end{equation}
Now, we choose the decays with FCNC that have contributions from the
textures matrices in Table \ref{tab:matrices}. The channels listed in
Table \ref{tab:4} were chosen because their branching ratios have been measured with  specific experimental errors allowing us to find a clear-cut valid range for the free parameters.

\begin{table}[ph]
 \tbl{Experimental measurements of the branching ratios for pseudoscalar neutral meson decays with flavor violation\cite{Olive2016,ATLAS-CONF-2018-046}.}{
    \begin{tabular}{l l}%
      \hline\hline
      \textbf{Channel}&\textbf{Branching Ratio}\\\hline
      $K_L^0 \rightarrow \mu^+\mu^-$& $6.84\pm 0.11 \times10^{-9}$ \\
			
      $K_L^0 \rightarrow e^+e^-$& $9\pm^6_4 \times10^{-12}$ \\
		
      $B^0 \rightarrow \mu^+\mu^-$& $1.6\pm 1.6 \times10^{-10}$\\
		
      $B_s^0 \rightarrow \mu^+\mu^-$& $2.8\pm ^{0.8}_{0.7} \times10^{-9}$ \\\hline\hline	
    \end{tabular}

    \label{tab:4}}
\end{table}

In order to obtain the allowed regions for $\tilde\chi^q_{2,ij}$ and $c$'s,
we perform a simple numerical analysis. We build the function
\begin{equation}
  \label{eq:chi_squared}
  \tilde \xi^2(\tan\beta,M_A,\chi_{2,ij}^{d,\ell}) =\frac{1}{N} \sum_{a=1}^N\frac{\left(\mathcal{B}^{\text{exp}}_a-\mathcal{B}^{\text{th}}_a\right)^2}{\left(\Delta\mathcal{B}^{\text{exp}}_a\right)^2},
\end{equation}
where $N$ is the number of channels with neutral flavor violation and
the theoretical branching ratio
$\mathcal B^{\text{th}}_a(\tan\beta,M_A,\chi_{2,ij}^{d,\ell})$
involves a minimal set of $\chi_{2,ij}^{d,\ell}$ couplings. The regions are obtained by using the restriction
$\tilde \xi^2\leq 1$.

The allowed region obtained from (\ref{eq:chi_squared}) depends on the order of magnitude of the $c$'s parameters. We estimate the range of values from neutral meson mixing bounds. Due to our analysis is only significative when it is posible to separate the long distance from short distance contributions, we extract the order of magnitude for the $c$'s parameters from the $B_{s,d}^0-\bar B^0_{s,d}$ mixing, where long distance contributions are very small, because b- and d-quarks masses are small compared with the electroweak scale and the top quark mass\cite{Urban1998}. The most general effective Hamiltonian for $\Delta B=2$ processes we can be writen as \cite{Becirevic2002}
\begin{equation}
  \label{eq:eff_hamiltonian}
  \mathcal H_{\text{eff}}^{\Delta B=2}=\sum_{q=s,d}\left(\sum_{i=1}^5C^q_i\mathcal O_i^q+\sum_{i=1}^3\widetilde C^q_i \widetilde{\mathcal O}^q_i\right),
\end{equation}
with
\begin{equation}
  \label{eq:operadores2}
  \begin{split}
    \mathcal O_1^q=&   (\bar b^\alpha\gamma_\mu P_L q^\alpha)(\bar b^\beta\gamma^\mu P_Lq^\beta),\\
    \mathcal O_2^q=&   (\bar b^\alpha P_Lq^\alpha)(\bar b^\beta P_Lq^\beta),\\
    \mathcal O_3^q=&   (\bar b^\alpha P_Lq^\beta)(\bar b^\beta P_Lq^\alpha),\\
    \mathcal O_4^q=&   (\bar b^\alpha P_Lq^\alpha)(\bar b^\beta P_Rq^\beta),\\
    \mathcal O_5^q=&   (\bar b^\alpha P_Lq^\beta)(\bar b^\beta P_Lq^\alpha),\\
  \end{split}
\end{equation}
where $\alpha$ and $\beta$ are color indices. The 6-dimensional operators $\widetilde{\mathcal O}^q_i$ are obtained from expressions (\ref{eq:operadores2}) by the exchange $P_L\to P_R$. Nevertheless, due to parity invariance in QCD, these operators lead to the same matrix elements than those without tilde\cite{Aoki2019}.
Thus, the mass splitting has two contributions written as
\begin{equation}
  \label{eq:mass_spliting}
  \Delta M_{q}=\Delta M^\text{SM}_{q}+\Delta M^{NP}_{q}.
\end{equation}

\begin{figure}[ht!]
  \centering\begin{subfigure}[]{
        \includegraphics[scale=0.75]{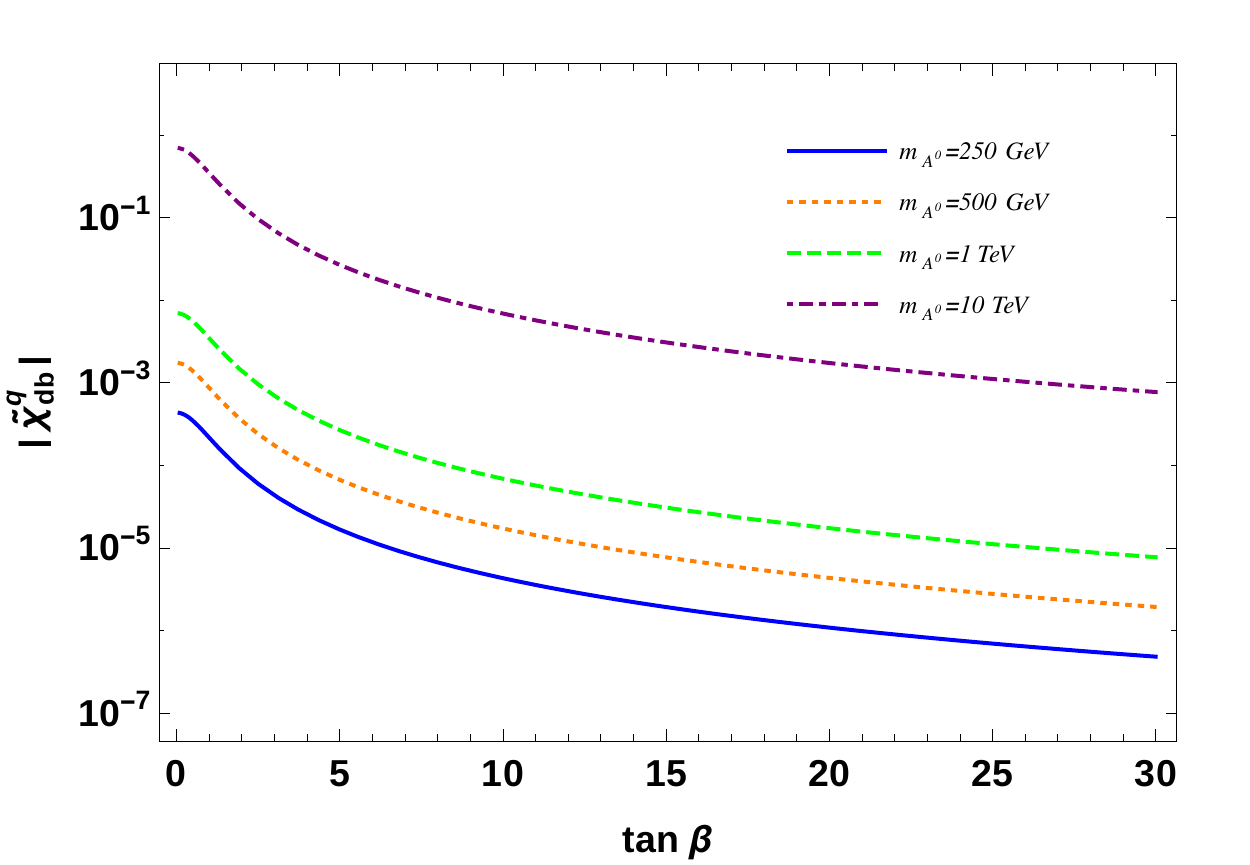}
        }
    \end{subfigure}
    \begin{subfigure}[]{
       \includegraphics[scale=0.75]{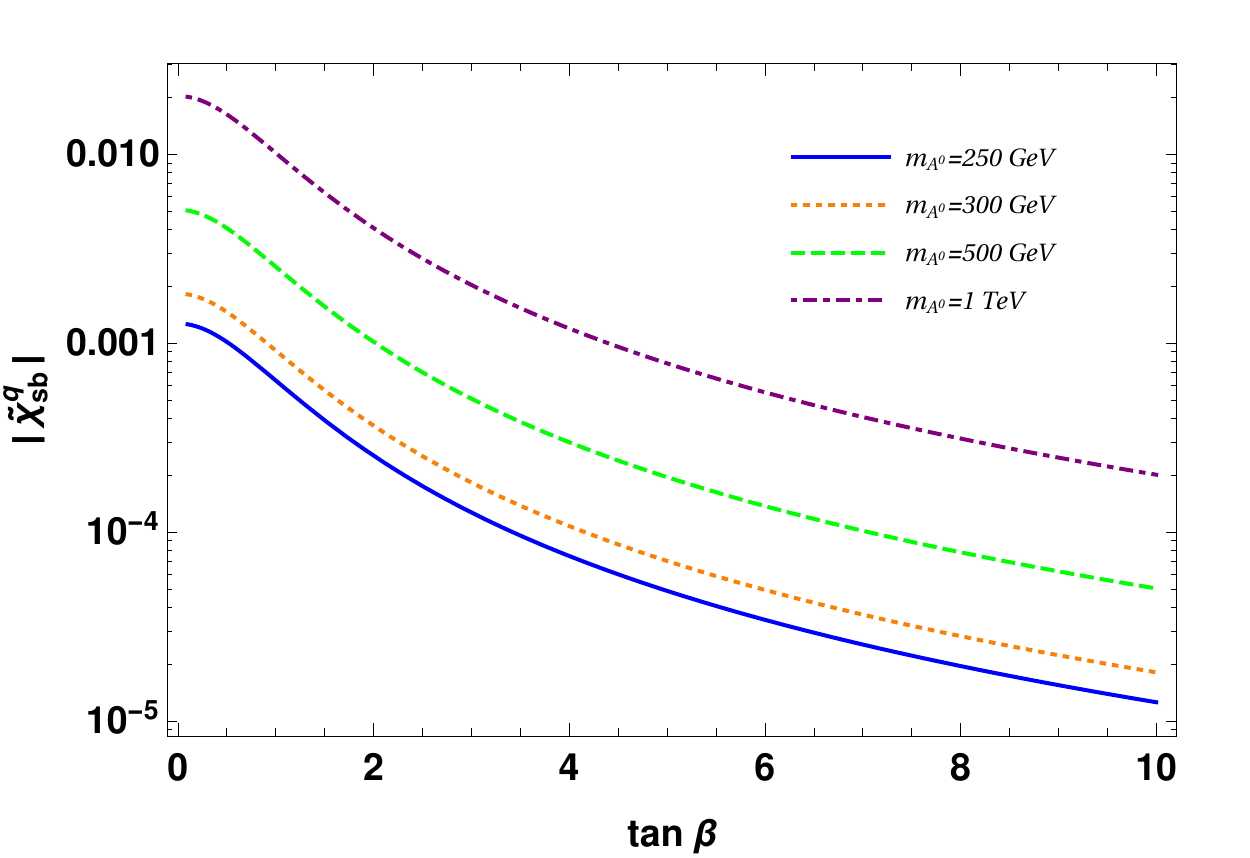}
        }      
    \end{subfigure}
    \caption{Here, we show the behavior of the bound on $|\tilde \chi^d_{2,bd}|$, that comes from (a) $B_d⁰-\bar B_d^0$ mixing in terms of $\tan\beta$.  Also we show the corresponding bound on $|\tilde \chi^d_{2,sb}|$ from (b) $B_s⁰-\bar B_s^0$ mixing in terms of $\tan\beta$. We observe how the the bound decrese for bigger values of $\tan\beta$. The optimistic bound for $M_{A^0}\sim 250$GeV and for small values of $\tan\beta$ is of the order of $\sim\mathcal O( 10^{-3})$}
    \label{mixBd_Chi_vs_tanb}
\end{figure}

\begin{figure}[ht!]
  \centering\begin{subfigure}[]{
        \includegraphics[scale=0.75]{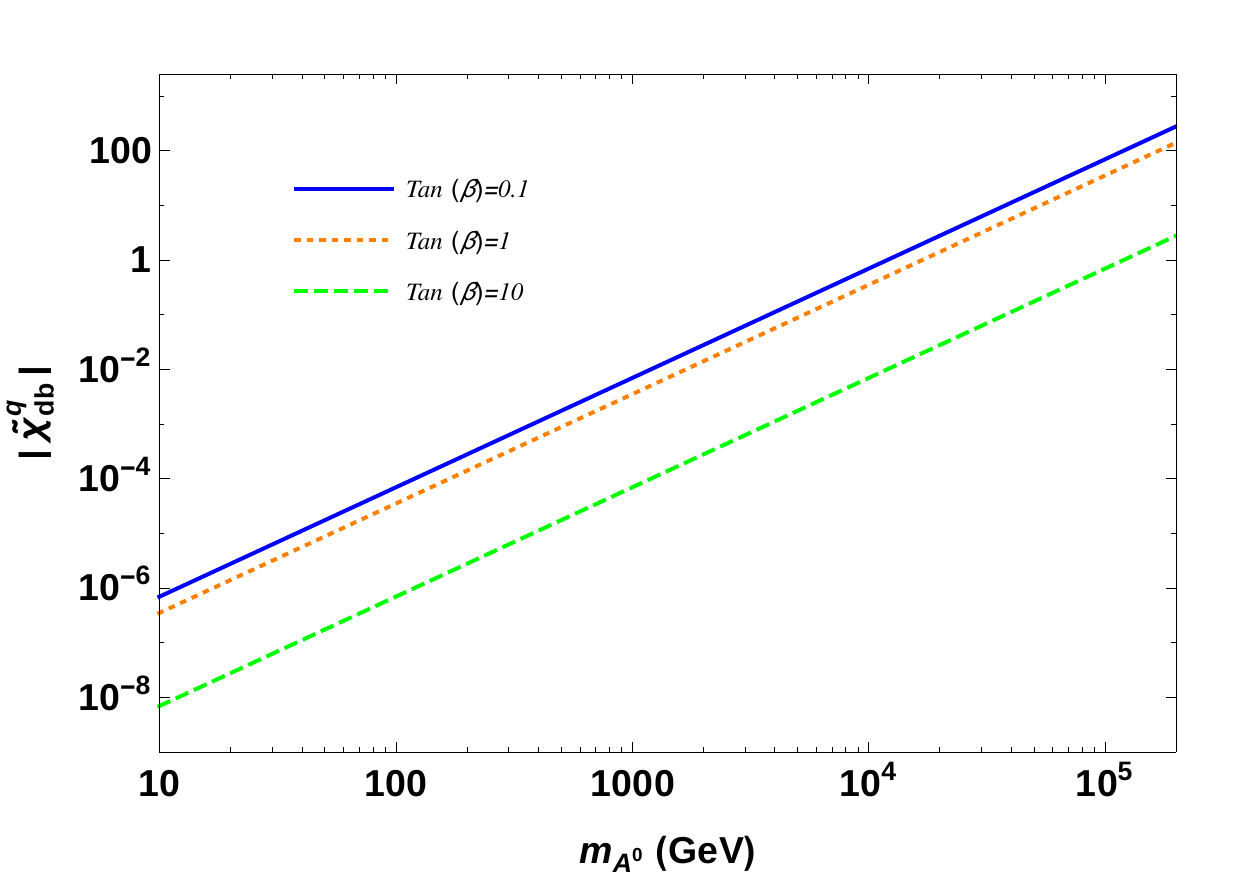}
        }
    \end{subfigure}
    \begin{subfigure}[]{
       \includegraphics[scale=0.75]{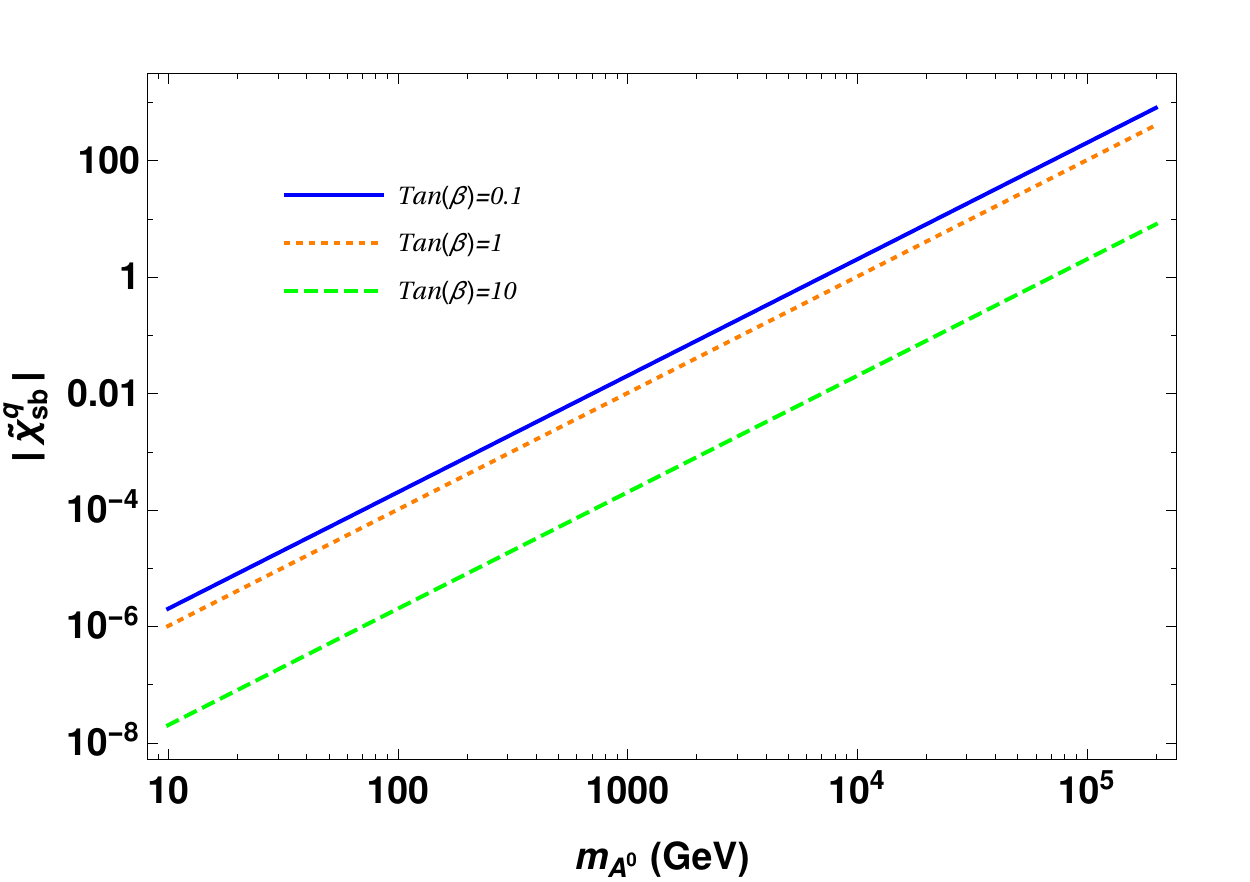}
        }      
    \end{subfigure}
\caption{Here, we show the behavior of the bound on $|\tilde \chi^d_{2,bd}|$, that comes from (a) $B_d⁰-\bar B_d^0$ in terms of the mass of the pseudoscalar $M_{A^0}$. Also we show the corresponding bound on $|\tilde \chi^d_{2,sb}|$ from (b) $B_s⁰-\bar B_s^0$ mixing in terms of the pseudoscalar masss $M_A^0$. In both cases the upper bound on the $\tilde\chi$'s parameters are of the order $\mathcal O(10^{-4})$ for $M_{A^0}\sim 250$GeV.}
\label{mixBd_Chi_vs_MA}
\end{figure}

The SM contribution to $B_{s,d}^0-\bar B^0_{s,d}$ mixing is given only by the operator $(\mathcal O^q_1)_R(m_b)$ renormalized with a widely known expression in terms of the bag parameter for the Vaccum Insertion Approximation \cite{Grossman1997}  and at the scale $\mu=m_b$
\begin{equation}
  \label{eq:matrix_elements}
  \Delta M^\text{SM}_{q}(m_b)=\frac{G_FM_W^2m_{B_q}}{6\pi^2}|\lambda_{tq}|^2S_0(x_t)\eta_{2,B}f_{B_q}^2\hat B_{B_q},
\end{equation}
where $S_0(x_t)$ is the Inami-Lim function and $\lambda_{tq}=V_{tq}V^*_{tb}$.
To parameterize $\Delta M_q^\text{NP}$ contribution we follow D. Be\'cirevi\'c, et. al\cite{Becirevic2002,Becirevic2002a,Huang2005}, where the renormalized matrix elements of the operators (\ref{eq:operadores2}) hves been calculated. We are interested in the pseudoscalar contribution at tree level, and up to this order we obtain
\begin{equation}
  \label{eq:deltamass_NP}
  \Delta M_q^\text{NP}(\mu)=|C_\text{NP}|\frac{M_{B_q}^2f_{B_q}^2}{\left[m_b(\mu)+m_q(\mu)\right]^2}\sigma(\mu),
\end{equation}
where we have define $\mathcal \sigma(\mu)=|-\frac{5}{24}B_2(\mu)+\frac{1}{24}B_3(\mu)+\frac{1}{24}B_4(\mu)+\frac{1}{12}B_5(\mu)|$ and $B_i(\mu)$ (for $i=2,3,4,5$) are the renormalized bag parameters at the scale $\mu$ ($\mu=m_b$). The coefficient $|C_\text{NP}|$ is calculated at the low energy limit of the generic 2HDM and have the form
\begin{equation}
  \label{eq:coefficient_NP}
  |C_\text{NP}|=\frac{1+\tan^2\beta}{2M_A^2}|\widetilde Y_{2,ij}^q|,
\end{equation}
where $\widetilde Y_{2,ij}^q$ is the corresponding Yukawa matrix element in the mass basis.
Thus the upper bound for $|\tilde\chi^d_{2,bq}|$ is given by
\begin{equation}
  \label{eq:up_bound}
  |\tilde \chi^d_{2,bq}|\leq\left(\frac{2M_A^2}{1+\tan^2\beta}\right)\frac{v(m_b+m_q)^2}{\sqrt{m_b m_q}M_{B_q}^2f_{B_q}^2}\frac{E_{B_q}}{\mathcal \sigma(m_b)}.
\end{equation}
The valid interval for $|\tilde \chi^d_{2,bq}|$ depends on the parameter $E_{B_q}=|\Delta M^\text{SM}-\Delta M^\text{Exp}|-\sqrt{\delta^2(\Delta M^\text{SM})+\delta^2(\Delta M^\text{Exp})}$, where we have added by quadrature the errors from the SM and the experiment. In figures (\ref{mixBd_Chi_vs_tanb}) and (\ref{mixBd_Chi_vs_MA})  we present the general updated behavior of this bound with respect to $\tan\beta$ and the pseudoscalar mass $M_A^0$. It is worth mentioning that this bound is independent of any particular assumption and can be used in any other version of the 2HDM with FCNC at tree level.


We find the maximum order of magnitude for the $c$'s parameters of the equations (\ref{eq:USpinChi}- \ref{eq:VSpinChi}) asking that, given certain pseudoscalar mass and a specific value for $\tan\beta$, the allowed region in the space $|\tilde\chi^d_{2,qb}|$ \emph{vs} $|\tilde\chi^\ell_{2,\mu\mu}|$ calculated from the channels $B_{s,d}\to\mu^+ \mu^-$, is below the bound given by the restriction (\ref{eq:up_bound}).

We want to point out that the assumption to use the same parameters $c$'s in the quark and leptonic sector allows us to introduce a light pseudoscalar mass because the bound for $|\tilde \chi^d_{2,bq}|$ ($q=s,d$) implies a bound on the leptonic sector and the combined effect renders small values for the factor $|\Gamma_{A^0q_iq_f}\Gamma_{A^0\mu \mu}|$ in the equation (\ref{Wilson_PCoeffcient}).  Also, if the order of magnitude for $|\tilde \chi^d_{2,bq}|$ become very small, it is an indication that the corresponding symmetry is not favored by the phenomenology.

\begin{figure}[ht]
  \centering
  \begin{subfigure}[]{
      \includegraphics[scale=4]{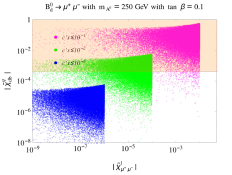}
    }
\end{subfigure}
      \begin{subfigure}[]{
  \includegraphics[scale=4]{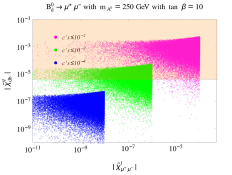}
      }
  \end{subfigure}
  \caption{In these plots, we show the allowed regions in the space $|\tilde\chi^d_{2,bd}|$ \emph{vs} $|\tilde\chi^\ell_{2,\mu\mu}|$ from the $B^0_d\to\mu^+\mu^-$ channel with a light pseudoscalar mass for different range of values for the parameters $c^U$'s and for (a) $\tan\beta=0.1$ and (b) $\tan\beta=10$. The shaded region is forbidden by the restriction (\ref{eq:up_bound}). The allowed parameters region completely contained below this bound is the one where $c^U,c^U_0,c_3^U\leq 10^{-3}$.} 
  \label{fig:allowed_reg_Bd}
\end{figure}

In figure \ref{fig:allowed_reg_Bd} we show, for a light pseudoscalar mass ($\sim 250$ GeV), the behavior of the allowed regions generated by the $B^0_d\to\mu^+\mu^-$ channel respect the order of magnitude for the $c$'s parameters. These regions are delimited on the above by the order of magnitude of the $c^U,c_0^U$ and $c_3^U$ parameters and on the right-hand side by the relation between the parameters $\chi^d_{2,db}$ and $|\chi^\ell_{2,\mu\mu}|$ (see Appendix \ref{appendix}), that is a consequence of our model. For the case of U-Spin, we choose $c^U_0,c^U_3,c^U\leq 10^{-3}$. Also, we note that the upper bound of $|\tilde\chi^\ell_{2,\mu\mu}|$ is always below the order of magnitude for $|\tilde\chi^d_{2,db}|$.

\begin{figure}[ht]
  \centering
  \begin{subfigure}[]{
  \includegraphics[scale=4]{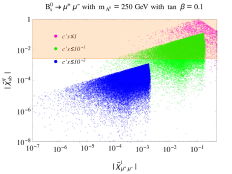}
      }
    \end{subfigure}
      \begin{subfigure}[]{
  \includegraphics[scale=4]{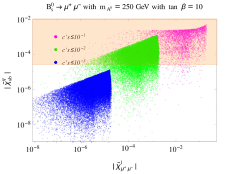}
      }
  \end{subfigure}
  \caption{Here, we show the allowed regions in the space $|\tilde\chi^d_{2,sb}|$ \emph{vs} $|\tilde\chi^\ell_{2,\mu\mu}|$ from the $B^0_s\to\mu^+\mu^-$ channel with a light pseudoscalar mass for different values for the parameters $c$'s and for (a) $\tan\beta=0.1$ and (b) $\tan\beta=10$. The shaded region is forbidden by the restriction (\ref{eq:up_bound}). The allowed parameters region completely contained below this bound is the one where $c^V,c^V_0,c_3^V\leq 10^{-2}$.} 
  \label{fig:allowed_reg_Bs}
\end{figure}

In figure (\ref{fig:allowed_reg_Bs}), we show a similar analysis for the V-Spin case. Here, the order of magnitude for $c^V, c_0^V$ and $c_3^V$ is of the order  $\sim \mathcal O(10^2)$ and it has a similar behavior than the former case. Nevertheless, the upper bound for $|\tilde\chi^d_{2,sb}|$ is below $|\tilde\chi^\ell_{2,\mu\mu}|$, unlike the previous case, because the hierarchy of masses of the involved fermions in V-Spin  is different respect the U-Spin case.
For the Iso-Spin case, we consider an optimistic order of magnitude for $c^I,c^I_0$ and $c^I_3$ taken from the U-Spin and V-Spin cases, that is $\sim\mathcal O(10^{-2})$. Now, we proceed to analyze every texture using the restrictions calculated before from equation (\ref{eq:chi_squared}) and the bound in (\ref{eq:up_bound}).

\subsection{V-Spin texture}
\label{sec:vspin}

\begin{figure}[h]
  \centering
  \includegraphics[scale=3]{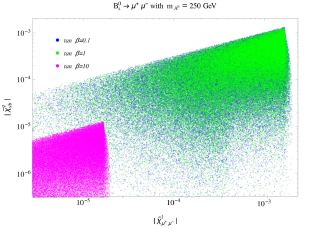}
  \caption{Parameter space for the V-Spin texture in a light pseudoscalar mass scenario ($M_{A^0}=250$ GeV). In this texture, the only FCNC contribution for leptonic meson decay is on $B^0_s\to\mu^+\mu^-$. The plotted regions were obtained for $\tan\beta= 0.1,1,10$. The region for values $\tan\beta\geq 1$ are not sensitive to this parameter.}
  \label{fig:VSpin250}
 \end{figure}

\begin{figure}[h]
  \centering
  \includegraphics[scale=3]{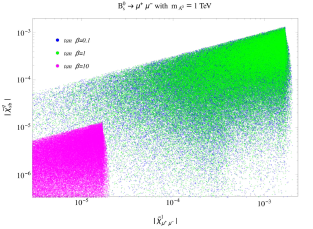}
  \caption{Parameter space for the V-Spin texture with a pseudoscalar mass $M_{A^0}=1$ TeV. In the V-Spin texture, the only FCNC contribution for leptonic meson decay is on $B^0_s\to\mu^+\mu^-$. Comparing this plot with figure (\ref{fig:VSpin250}), we note that for masses $\leq 1$ TeV, the dependence on $M_{A^0}$ is weaker.}
\label{fig:VSpin1000}
\end{figure}

\begin{figure}[h]
  \centering
  \includegraphics[scale=3]{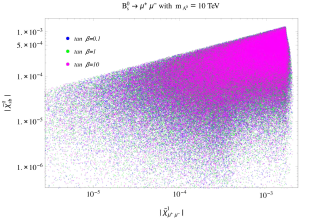}
  \caption{Parameter space for the V-Spin texture with a heavy pseudoscalar mass $M_{A^0}=10$ TeV. In this texture, the only FCNC contribution for leptonic meson decay is on $B^0_s\to\mu^+\mu^-$. For pseudoscalar masses of this scale, the parameter $\tan\beta$ is not important for values $\leq 10$.}
  \label{fig:VSpin10000}
\end{figure}

In the V-Spin texture, we only have contributions coming from the coupling between second and third generation, i. e. the NP contribution is on the $B_s\to \mu^+\mu^-$ channel. We found, as it was shown in the previous section, that the upper bound of $|\tilde\chi_{2,sb}^d|$ displays a behavior, with respect to $|\tilde\chi_{2,\mu\mu}^\ell|$, that can be explained in two parts
once $\tan\beta$ and the mass $M_A$ are fixed. For small values of $|\tilde\chi_{2,\mu\mu}^\ell|$, the parameter $|\tilde\chi_{2,sb}^d|$ increases to a maximum value, then it decreases monotonically as shown in figure (\ref{fig:VSpin250}) for a light pseudoscalar mass and in figures (\ref{fig:VSpin1000}) and (\ref{fig:VSpin10000}) for the typical heavy pseudoscalar masses, i. e. $1$ TeV and $10$ TeV respectively.

As it is expected, bigger regions are obtained for small values of $\tan\beta$ and for masses of the order $M_{A^0}\sim\mathcal O(10 \text{TeV})$ and the regions are essentially independent of $\tan\beta$. 


This maximum value for $|\tilde\chi_{2,sb}^d|$, that depends on the
chosen values for $\tan\beta$ and $M_{A^0}$, allows us to compare this prediction with previous analysis in versions with NFC and 2HDM-III.

\subsection{U-Spin texture}
\label{sec:uspin}
\begin{figure}[ht]
  \centering
  \includegraphics[scale=3]{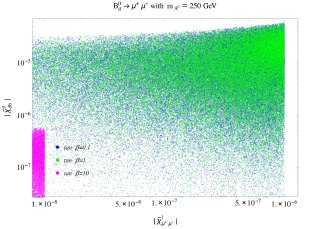}
  \caption{Parameter space for the U-Spin texture in a light pseudoscalar mass scenario ($M_{A^0}=250$ GeV). In this texture, the only FCNC contribution for leptonic meson decay is on $B^0_d\to\mu^+\mu^-$. The plotted regions were obtained for $\tan\beta= 0.1,1,10$. The region for values $\tan\beta\geq 10$ is very small, thus a scenario with big $\tan\beta$ and light pseudoscalar mass is no favored. In general, for this case, the complete region is small compared with the other cases.}
  
  \label{fig:USpin250}
 \end{figure}

\begin{figure}[ht]
  \centering
  \includegraphics[scale=3]{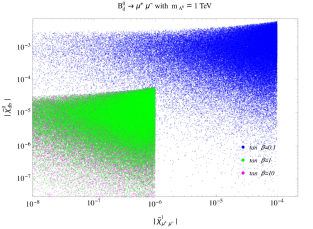}
  \caption{Parameter space for the U-Spin texture with a pseudoscalar mass $M_{A^0}=1$ TeV. In the U-Spin texture, the only FCNC contribution for leptonic meson decay is on $B^0_d\to\mu^+\mu^-$. Comparing this plot with figure (\ref{fig:USpin10000}), we note that for masses $\geq 1$ TeV, the dependence on $M_{A^0}$ is weaker.}
\label{fig:USpin1000}
\end{figure}

\begin{figure}[ht]
  \centering
  \includegraphics[scale=3]{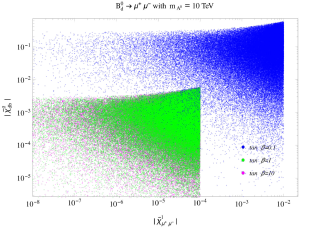}
  \caption{Parameter space for the U-Spin texture with a heavy pseudoscalar mass $M_{A^0}=10$ TeV. In this texture, the only FCNC contribution for leptonic meson decay is on $B^0_d\to\mu^+\mu^-$. For pseudoscalar masses of this scale, the bigger region is obtained for $\tan\beta\simeq 0.1$}
  \label{fig:USpin10000}
\end{figure}

In the U-Spin texture, the contribution is on $B^0_d\to\mu^+\mu^-$. This case is characterized by the fact that the coupling depends on the ratio of highly hierarchical masses, where we have obtained
\begin{equation}
  \label{eq:chi13}
|\tilde\chi_{2,13}^q|=|c^U|\left|\sqrt{\frac{m_d}{m_b}}(c^U_0+c^U_3)+\sqrt{\frac{m_b}{m_d}}(c^U_0-c^U_3)\right|,
\end{equation}
thus the parameter $|\tilde\chi^q_{2,13}|$ is dominated by the first term when $c^U_0\simeq c_3^U$ and it is very small due to the factor $\sqrt{\frac{m_d}{m_b}}$.
The behavior is shown in Figures (\ref{fig:USpin250}-\ref{fig:USpin10000}).


In general, this channel constrains more NP contributions with a light pseudoscalar mass. This results means that U-Spin is not favored by the phenomenology in this scenario. Nevertheless, as it is noticed in figure (\ref{fig:USpin10000}), the upper bound on $|\tilde\chi^d_{2,db}|$ can be of the order $\sim\mathcal{O}(1)$ for $M_{A^0}\sim 10$ TeV. This is not a new result, but it coincides with analysis done by other authors in generic versions of the 2HDM. Our model shed an upper bound for $|\tilde\chi^\ell_{2,\mu\mu}|$ of the order $\sim\mathcal O(10^{-2})$ for this texture.
 
\subsection{Iso-Spin Texture}
\label{sec:isospin}
\begin{figure}[ht]
  \centering
  \includegraphics[scale=3]{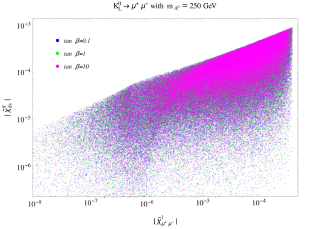}
  \caption{Parameter space for the Iso-Spin texture with a light pseudoscalar mass scenario ($M_{A^0}=250$ GeV). In this texture, we use only the short distance contribution to $K^0_L\to\mu^+\mu^-$ in order to bound $|\tilde\chi^d_{2,ds}|$. The plotted regions were obtained for $\tan\beta= 0.1,1,10$. The regions are not significantly dependent on $\tan\beta$. In this case, the order of magnitude of the upper bound for $|\tilde\chi^d_{2,ds}|$ is always below the one for $|\tilde\chi^\ell_{2,\mu\mu}|$ .}
 \label{fig:ISpin250}
 \end{figure}

\begin{figure}[ht]
  \centering
  \includegraphics[scale=3]{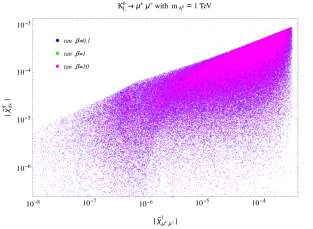}
  \caption{Parameter space for the Iso-Spin texture with a pseudoscalar mass $M_{A^0}=1$ TeV. Comparing this plot with figure (\ref{fig:ISpin250}), we see that both scenarios are similar for light pseudoscalar masses up to $M_{A^0}\sim 1$ TeV.}
\label{fig:ISpin1000}
\end{figure}

\begin{figure}[ht]
  \centering
  \includegraphics[scale=3]{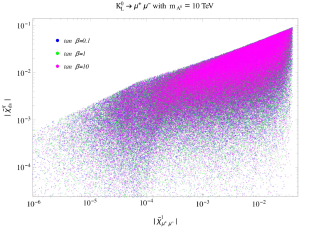}
  \caption{Here, we show the parameter space for the Iso-Spin texture with a heavy pseudoscalar mass $M_{A^0}=10$ TeV. Only the short distance contribution were taken into account to bound $|\tilde\chi^q_{2,ds}|$ and $|\tilde\chi^\ell_{2,\mu\mu}|$. The dependence in $\tan\beta$ does not change the size of the region, nevertheless, this case renders the bigger allowed region compared with the cases where $M_{A^0}\leq 1$ TeV.}
  \label{fig:ISpin10000}
\end{figure}

In this texture, the first and second generations are mixed and the channels that have NP contributions are $K_L\to\mu^+\mu^-$ and $K_L\to e^+e^-$. The form factors uncertainties are relevant for this particular decay. We will focus in the $K_L\to\mu^+\mu^-$ channel because it has the smallest theoretical uncertainty\cite{Gorbahn2006}. The proccess with 2 electrons in the final state provides no new information about the $|\tilde\chi^q_{sb}|$ limits. In order to analyze the case of Iso-Spin, we need to take into account the different contributions to $K_L\to\mu^+\mu^-$. In this case the branching ratio can be decomposed in two contributions
\begin{equation}
  \label{eq:branching_ratio}
  B(K_L\to\mu^+\mu^-)=|\mathcal Re A_{\text{LD}}+\mathcal Re A_{\text{SD}}|^2+|\mathcal Im A|^2
\end{equation}
The first term in (\ref{eq:branching_ratio}) is called the dispersive contribution to the decay and contains electroweak FCNC up to one-loop SM diagrams and possible NP contributions. Also, there is a contribution coming from process with virtual photons that can be dismissed. In order to use this decay to bound NP parameters, we need to calculate the second term. This last term is called the absortive part and it is controlled in a model independent way by the branching ratio of  $K_L\to\gamma\gamma$ process.This contribution, in the unitary gauge, can be calculated as follows\cite{Eeg1998,DAmbrosio1998}
\begin{equation}
  \label{eq:absortive_part}
  |\mathcal Im A|^2=\frac{\alpha_\text{EM}^2 m_\mu^2}{2 M_K^2\beta_{\mu}}\left[\ln{\frac{1-\beta_\mu}{1+\beta_\mu}}\right]^2\text{BR}(K_L\to\gamma\gamma),
\end{equation}
where $\beta_\mu=\sqrt{1-\frac{4m_\mu^2}{M_K^2}}$. Using updated experimental results for $\text{BR}(K_L\to\gamma\gamma)$, we obtain that the absorptive contribution is given by
\begin{equation}
  \label{eq:BR_abs}
  |\mathcal Im A|^2=(6.53\pm 0.04)\times10^{-9}.
\end{equation}
The experimental value\cite{Olive2016} $ BR(K_L\to\mu^+\mu^-)=(6.84\pm 0.11)\times 10^{-9}$, leads to the dispersive part
\begin{equation}
  \label{eq:dispersive_part1}
  |\mathcal Re A_{\text{LD}}+\mathcal Re A_{\text{SD}}|^2=(3.03\pm 1.20)\times10^{-10},
\end{equation}
where we have added the errors by quadrature. Using this result, we can bound $|\tilde\chi_{2,bd}^d|$ because the coupling with the pseudoscalar is part of the short distance contribution.
In figures (\ref{fig:ISpin250}), (\ref{fig:ISpin1000}) and (\ref{fig:ISpin10000}), we show the result for this texture. For pseudoscalar masses below $1$ TeV, there is no significant dependence on $\tan\beta$ and the bounds. On the other hand, as it is expected, for masses of the order $10$ TeV, the allowed region gives an upper bound to $10^{-1}$ whether for $|\tilde\chi^d_{2,ds}|$ and for $|\tilde\chi^\ell_{2,\mu\mu}|$.


\section{Conclusions}\label{conclusions}
 This work is summarized as follow
\begin{itemize}
\item We have taken the parametrization of the PA-2HDM to introduce Yukawa matrices generated through flavor transformations belonging to subgroups of $SU(3)$. These subgroups are Iso-Spin, U-Spin and V-Spin. With this restricted set of transformations, we generate couplings that mix pairwise the generations of fermions. In this context, we analized the phenomenology of leptonic decay of the mesons $B^0_{s,d}$ and $K^0_L$ going to a pair of muons.
\item Also, we reach a severe reduction of free parameters compared with the 2HDM-III to describe small FCNC, introducing a parallel texture between quarks and leptons. This assumption leads, not only to a scenario similar to a universal texture, but also to a relation between Yukawa couplings in the quark sector with the couplings in the leptonic sector. This was done using the same parameters of the flavor transformations in both sectors. The order of magnitude for the parameters were determined using the bounds impose by $B^0_s-\bar B^0_{s}$ and $B^0_d-\bar B^0_{d}$ mixing.
\item For a relatively light mass for the pseudoscalar $A^0$ ($m_A\sim 250$GeV), the imposed flavor symmetry generates very small FCNC and in some cases null. The small upper bounds obtained for $|\tilde\chi^d_{2,qb}|$ ($q=s,d$) from $B^0_{s,d}$ mixing, affects the upper bound on $|\tilde\chi^\ell_{2,\mu\mu}|$. The combined effect of these bounds is the suppression of the branching ratios of $B^0_s\to\mu^-\mu^+$ and $B_d^0\to\mu^-\mu^+$. We analized the cases with masses of $1$ TeV and $10$ TeV for values of $\tan\beta$ where such bounds significantly change.
  \item Updated values for the decay constant $f_K$ and the branchig ratio for $K_L^0\to\gamma\gamma$ have been used to calculate the short distance contributions to $K_L^0\to \mu^+\mu^-$. We have obtained a bound to constrain the short distance contribution, coming from the exchange of the pseudoscalar of the 2HDM. With this value, we have delimited numerically $|\tilde\chi^d_{2,sd}|$ and $|\tilde{\chi}^\ell_{2,\mu\mu}|$ for the Iso-Spin texture, assuming that the parameters of the flavor transformations of Iso-Spin, i. e. $c^I, c^I_0$ and $c_3^I$, are of the same order of magnitude than in the cases of V-Spin and U-Spin.
\end{itemize}

In this analysis, we have demostrated that a light pseudoscalar mass ($M_{A^0}\sim 250$ GeV) with a flavor symmetry, parameterized by restricted flavor transformations of $SU(3)$, can generate small FCNC. Also, a parallel structure of Yukawa matrices between quarks and leptons could explain the suppressions but it is not the only scenario. The order of magnitude of the upper bounds obtained for $|\tilde\chi^d_{2,ij}|$ and $|\tilde\chi^\ell_{2,\mu\mu}|$ are competitive with those found in other analysis in the context of the 2HDM-III. The results showed in figures (\ref{fig:VSpin250}-\ref{fig:ISpin10000}) are not enough to determined if one of the specific Yukawa texture analized in this work, is favored by the phenomenology. Nevertheless, the relevant result of this work is the introduction of a parametrization that bounds leptonic flavor violating processes from FCNC in the quark sector. Other scenarios and more observables must be analized in order to restrict more the regions shown in previous sections.

As we mention before, only the $B_s\to \mu^{-}\mu^{+}$ channel is sensitive to the V-spin contributions, the channel $B_d\to \mu^{-}\mu^{+}$ to U-spin, and the channel $K_L \to \mu^{-}\mu^{+}$ to Iso-spin. The texture leads to FCNC in agreement with the experimental measurements with only 3 parameters for every case. In particular, for the case of parallel textures between down-type quarks and leptons, where parallel textures mean that both have the same pattern of zeros, the parameters $\tilde\chi$'s take simple forms weighted by the square root of the quark masses ratio involved in the corresponding decay.

In general, we have demostrated that the upper bound on the $|\tilde\chi_{2,ij}^{q}|$ have a behavior regulated by 3 free parameters ($c$'s), the fermion mass ratios, the pseudoscalar mass $m_A$ and $\tan\beta$. For the $B_{s,d}$ and $K_L$ channels, $|\tilde\chi_{2,ij}^{q}|$ rises for small values of $|\tilde\chi_{\mu \mu}^{\ell}|$, attaining a maximum value and decreasing thereafter. 

Also, we found that if a pseudoscalar with a relatively small mass and an extra $U(1)$ flavor symmetry exist, the branching ratios of leptonic meson decays might not be sensitive to the NP with the current experimental precision.

\section*{Acknowledgments}

ACM is supported by the Mexican National Council for Science and
Technology (CONACyT) scholarship scheme and SGA is supported by
PRODEP (Mexico).

\section*{Appendix A}\label{appendix}
We have used the same values for parameters $c$'s to calculate the $|\tilde\chi_{2,ij}^d|$ and $\tilde\chi^\ell_{2,ij}$. Through the expressions (\ref{eq:ISpinChi}-\ref{eq:VSpinChi}), we can obtain relations between the $|\chi^f_{2,ij}|$ parameters. Here, we show the explicit expressions for all the cases studied before. These expressions allow us to bound the $|\tilde\chi^\ell_{2,\mu\mu}|$ as it was describe in section \ref{section3}. In order to simplify the notation, we have omitted the index in the masses that indicates the corresponding sector. 

\subsection{Iso-Spin}
\label{eq:App_IsoSpin}
\begin{eqnarray}
  c_0^I&=&\sqrt{\tilde \chi_{2,33}^f}\label{eq_c0}\\
  c_ 3^I&=&\frac{(m_1^2+m_2^2)\sqrt{\widetilde\chi_{2,33}^f}}{m_2^2-m_1^2}\nonumber\\
       &&\qquad\qquad\pm\frac{\sqrt{m_1^4\widetilde\chi_{2,11}^f+m_2^4 \widetilde\chi_{2,22}^f-m_1^2m_2^2(\widetilde\chi_{2,11}^f+\widetilde\chi_{2,22}^f-4\widetilde\chi_{2,33}^f)}}{m_2^2-m_1^2}  \label{eq_c3} \\
  c^I&=&\frac{\sqrt{m_1m_2}}{m_2^2-m_1^2}\left[(m_2^2-m_1^2)(\widetilde\chi_{2,11}^f-\widetilde\chi_{2,11}^f)-4(m_2^2+m_1^2)\widetilde\chi_{2,33}^f\right.\nonumber\\
&&\qquad\left.\pm 2\sqrt{\widetilde\chi_{2,33}^f}\sqrt{m_1^4\widetilde\chi_{2,11}^f+m_2^4\widetilde\chi_{2,22}^f-m_1^2m_2^2(\widetilde\chi_{2,11}^f+\widetilde\chi_{2,22}^f-4\widetilde\chi_{2,33}^f)}\right]^{1/2}\label{eq_c}
\end{eqnarray}

\subsection{U-Spin and V-Spin}
To obtain the expresions for the U-Spin case, we have to exchange $\widetilde\chi_{2,33}^f\to\widetilde\chi_{2,22}^f$ and $m_2\to m_3$ in equations (\ref{eq_c0}-\ref{eq_c}). And for the V-Spin case, the exchange $\widetilde\chi_{2,11}^f\to \widetilde\chi_{2,22}^f$ and $m_1\to m_2$ from the corresponding expressions for U-Spin.

\bibliography{paper_IJPA_to_arXiv}

\end{document}